\documentstyle[12pt]{article}
\newcommand{\ald}{\dot \alpha}
\newcommand{\bed}{\dot \beta}

\newcommand{\ej}{\cal E}
\newcommand{\hej}{\hat {\ej}}
\newcommand{\sej}{{\cal S}}
\newcommand{\shej}{\widehat {\sej}}
\newcommand{\th}{\theta}
\newcommand{\bth}{\bar \theta}
\newcommand{\beq}{\begin{equation}}
\newcommand{\eeq}{\end{equation}}

\begin{document}
\begin{titlepage}
\begin{center}

\huge{\bf On tree form-factors in (supersymmetric) Yang-Mills theory}

\vspace{1.5cm}

\large{\bf K.G.Selivanov }

\vspace{1.0cm}

{ITEP,B.Cheremushkinskaya 25, Moscow,117259, Russia\\
 email: selivano@heron.itep.ru}

\vspace{1.9cm}

{ITEP-TH-47/98}

\vspace{1.0cm}

\end{center}
              

\begin{abstract}
{\it Perturbiner}, that is, the solution of field equations which is a 
generating function for tree form-factors in $N=3$ $(N=4)$ supersymmetric
Yang-Mills theory, is studied in the framework of twistor formulation of the
$N=3$ superfield equations. In the case, when all one-particle asymptotic 
states belong to the same type of $N=3$ supermultiplets (without any 
restriction on kinematics), the solution is described very explicitly.
It happens to be a natural supersymmetrization of the self-dual
perturbiner in non-supersymmetric Yang-Mills theory, designed to describe the 
Parke-Taylor amplitudes. In the general case,
we reduce the problem to a neatly formulated algebraic geometry problem
(see Eqs(\ref{5.15i}),(\ref{5.15ii}),(\ref{5.15iii})) and propose an 
iterative algorithm for solving it, however we have not been able to find a 
closed-form solution. Solution of this problem would, of course, produce a 
description of all tree form-factors in non-supersymmetric Yang-Mills 
theory as well. In this context, the $N=3$ superfield formalism may be 
considered as a convenient way to describe a solution of the 
non-supersymmetric Yang-Mills theory, very much in the spirit 
of works by E.Witten \cite{Witten} and by  J.Isenberg,
P.B.Yasskin and P.S.Green  \cite{2}. 
\end{abstract}
\end{titlepage}

\newpage
\setcounter{equation}{0}

\section{Introduction}
It is well known that multi-particle amplitudes, even in the tree approximation, are generically out of reach by means of available field theoretical methods,
though, there is a lot of efforts and achievements in this direction 
(see \cite{mapa}-\cite{RS4} and other references to the paper
\cite{mapa}). In principle, the problem can be considered as a purely
technical one, since the $n$-particle tree amplitude is, according to the 
text-book rules, represented as a sum of a number of rational functions of 
momenta of the particle ( a term of the sum is a contribution of a Feynman 
diagram, 
and the amplitude is a sum of contributions of a number of diagrams). However, the number of terms growths enormously with the number of particles , so that 
the total expression becomes untreatable when
the number of particles becomes bigger than, say, 9, to say nothing about 
arbitrary $n$. On the other side, sometimes a final expression for the 
amplitude is essentially simpler than the intermediate ones. There are known 
cases when cancellations among contributions of different Feynman diagrams 
are just wonderful \cite{ParkeTaylor}-\cite{Argyres},
 \cite{LRB}. These cases are, essentially, scalar field 
amplitudes with most of the external particles at threshold 
\cite{Voloshin}-\cite{Argyres}, \cite{LRB}
and the so-called like-helicity amplitudes in Yang-Mills theory
(that is, amplitudes with most of the external gluons in the same helicity
state)\cite{ParkeTaylor}, \cite{BerendsGiele}. If one is optimistic concerning possible cancellations in more general cases, say, in the case of Yang-Mills 
amplitudes with arbitrary helicities, one should look for a way to avoid 
those intermediate steps. A possible idea is to use the classical field 
equations since the tree amplitudes can, of course, be obtained from a
classical solution of the field equations. This approach has been discussed 
in the classical text-books \cite{Faddeev}, \cite{Itzikson}, and it has 
recently resurrected in the literature. The threshold amplitudes in scalar 
field 
theories were obtained from spatially uniform classical solutions, thus, the 
field equations reduced to ordinary differential equations 
\cite{Voloshin}-\cite{Argyres}, \cite{LRB}.
 The like-helicity Yang-Mills amplitudes were related to 
solutions of the self-duality equations in \cite{Bardeen}, \cite{Selivanov1}.

From our subjective point of view, one of the most interesting by-products of 
the above developments
was the idea of {\it perturbiner}, or ptb-solution, \cite{RS1}-\cite{RS4}. 

To define the perturbiner, one
first fixes a solution of linearized field equations (which are assumed to 
describe asymptotic one-particle states)
of the type of
\beq
\label{1}
\phi^{(1)}=\sum_{J=1}^{L}a_{J}\epsilon_{J}t_{J}e^{k_{J}x}=
\sum_{J}^{L}\epsilon_{J}{\hej}
\eeq
where $x$ stands for a space-time coordinate, $k_{J}$ stands for a momentum
of the $J$-th particle, $\epsilon_{J}$ stands for a polarization of the 
$J$-th particle, $t_{J}$ stands for a ``polarization'' of the $J$-th
particle in the internal space (e.g. $t_{J}$ is a generator of the color 
group), $a_{J}$ is a symbol of annihilation/creation operator,
\beq
\label{harm}
{\hej}_{J}=t_{J}{\ej},\; {\ej}=a_{J}e^{k_{J}x}
\eeq
The perturbiner is a complex solution of the field equations which is 
a formal power series in the ``harmonics'' ${\ej}_{J}, J=1, \ldots, L$ Eq.
(\ref{harm}), the first order term of the 
series being just the solution Eq.(\ref{1}). 

Notice that $x$-dependence of the perturbiner comes only via monomials
in  ${\ej}_{J}, J=1, \ldots, L$, on which differential operators entering
the field equations act as algebraic operators, and existence/uniqueness 
\footnote{In gauge theories the uniqueness is, of course, modulo gauge
transformations} of the perturbiner normally takes place provided the set 
of momenta, $k_{J},  J=1, \ldots, L$, is non-resonant, that is, provided
none of linear 
combinations of $k_{J},  J=1, \ldots, L$ with positive integer
coefficients including 
more than one momentum
 gets to the mass shell. The physical 
meaning of the perturbiner is that its expansion in powers of symbols
 $a_{J}, J=1, \ldots, L$ generates tree form-factors in the theory, 
\beq
\label{2}
{\phi}^{ptb}(x, \{k\}, \{a\})=\sum_{l=1}^{L}\sum_{\{J\}}a_{J_{1}} \ldots 
a_{J_{l}}<k_{J_{1}}, \ldots , k_{J_{l}}|{\phi}(x)|0>_{tree}
\eeq

Notice that all the external mass-shell particles are arbitrarily considered
as the out- ones. In principle, those with negative frequency should be 
considered as out-
states while those with positive  frequency should be considered as in-states
but at tree level analytical continuation from negative to positive frequency 
is trivial, so we do not distinguish them. The form-factors
$<k_{J_{1}}, \ldots , k_{J_{l}}|{\phi}(x)|0>_{tree}$ in Eq.(\ref{2})
are in the coordinate representation. The monomials in the harmonics 
${\ej}_{J}, J=1, \ldots, L$
will produce the momentum conservation $\delta$-functions after transformation 
to the momentum space.

It is very convenient  to add to the above definition of the 
perturbiner the requirement of nilpotency of the symbols  
$a_{J}, J=1, \ldots, L$, that is 
\beq
\label{nilpo}
a_{J}^{2}=0.\footnote{This condition has 
nothing to do with the statistics of the particles considered.
Say, for bosons the symbols still commute, while for fermions -
anticommute.}
\eeq
 In terms of the form-factors the nilpotency means that 
form-factors with identical asymptotic states will not appear in the 
expansion of the perturbiner in powers of $a_{J}, J=1, \ldots, L$ 
(see Eq.(\ref{2})). Clearly, it does not assume any loss of generality if the 
perturbiner is known for an arbitrary number  $L$ of the asymptotic 
one-particle states, since the form-factors with 
identical asymptotic states can be obtained from those with all asymptotic 
states different. It is clear that under the nilpotency assumption the 
perturbiner
is, in fact, not the power series but just a polynomial in (nilpotent)
harmonics ${\ej}_{J}, J=1, \ldots, L$ Eq.(\ref{harm}). Actually, for massless
particles, the nilpotency assumption is necessary for the non-resonantness
condition, because any multiple of a light-like momentum is again a light-like
momentum, and the nilpotency makes these multiples irrelevant. 

Anywhere below we shall assume the nilpotency condition Eq.(\ref{nilpo}).

We note that textbooks (see, e.g., \cite{Faddeev}, \cite{Itzikson}) offer 
another 
definition of solution of field equations, generating tree amplitudes 
in the theory (they use the so-called Feynman asymptotic condition to define 
it). 

Our definition above proved to be very convenient, in particular, it happened
to be very conveniently compatible with the twistor description of solutions 
for the gauge self-duality equation (\cite{RS1}), for the gravitational 
self-duality
equations {\cite{RS2}) and for the gauge-gravitational self-duality
equation (\cite{RS3}). The traditional finite-action and reality conditions
are substituted in the case of perturbiner by the condition of analyticity
in the harmonics ${\ej}_{J}, J=1, \ldots, L$. Eqs.(\ref{1}),(\ref{harm}).
It is worth to explain that the perturbiner obeys the self-duality equations 
instead of the full equations, such as Yang-Mills equations or Einstein
equations, when all polarizations entering  Eq.(\ref{1}) describe the same 
helicity state. Self-dual perturbiner generates only like-helicity
form-factors. In this way the so-called Parke-Taylor
amplitudes \cite{ParkeTaylor}, \cite{BerendsGiele} are very unusually
described in terms of meromorphic functions on an auxiliary $CP^{1}$ space.

In this paper we describe perturbiner in $N=3$ $(N=4)$ supersymmetric 
 Yang-Mills theory.
$N=3$ supersymmetric Yang-Mills is equivalent to $N=4$ super Yang-Mills but
$N=3$ superfield formalism is more naturally combined with twistors 
\cite{Witten}, that is why we follow $N=3$
notation. $N=4$ super Yang-Mills multiplet consists of the following
particles:
\begin{eqnarray}
\label{N=4}
1 \times 1 \: {\rm (positive\, helicity\, gluon)}\nonumber\\
4 \times 1/2 \: {\rm (positive\, helicity\, gluinos)}\nonumber\\
6 \times 0 \: {\rm (scalars)}\nonumber\\
4 \times -1/2 \: {\rm (negative\, helicity\, gluinos)}\nonumber\\ 
1 \times -1 \: {\rm (negative\, helicity\, gluon)}
\end{eqnarray}
This multiplet decomposes into two $N=3$ multiplets as follows
\begin{eqnarray}
\label{N=3a}
1 \times 1\nonumber\\
3 \times 1/2 \nonumber\\
3 \times 0 \nonumber\\
1 \times -1/2 
\end{eqnarray}
\begin{eqnarray}
\label{N=3b}
1 \times 1/2 \nonumber\\
3 \times 0  \nonumber\\
3 \times -1/2 \nonumber\\ 
1 \times -1
\end{eqnarray}
It occurs that if one includes only states from one type of the $N=3$ 
multiplets, say,
the one from the table (\ref{N=3a}) (in arbitrary kinematics),
that is all plane waves in Eq.(\ref{1}) belong to the same type of the $N=3$ 
multiplets
(with arbitrary momenta  $k_{J},  J=1, \ldots, L$), the solution is obtained 
from the (non-supersymmetric)
self-dual perturbiner \cite{RS1} by substituting the harmonics ${\ej}$ 
Eq.(\ref{harm}) with their supersymmetric extensions ${\sej}$ Eq.(\ref{sharm}).
Such solution will be called {chiral} $N=3$ perturbiner.
If one includes both types of the $N=3$ multiplets 
the problem
becomes much more complicated.\footnote{ This is not surprising because all 
tree
form-factors in non-supersymmetric Yang-Mills are contained among the  $N=4$
form-factors (when all external particles are gluons, the fermions and the 
charged scalars can appear only in loops). Actually, all $N=3$ machinery
can be viewed on as a convenient way to describe solutions of ordinary
Yang-Mills equations, very much in the spirit of \cite{Witten} 
(see, also, \cite{2}).} In this case we describe a twistor reformulation
of the problem, we show how it can be solved iteratively, but we have not been 
able to find a closed-form solution of the problem. 
Nevertheless, the problem is reduced to a 
neatly formulated algebraic geometry problem, Eqs.(\ref{5.15i}),
(\ref{5.15ii}),(\ref{5.15iii}),  
and we feel that the complete solution might be somewhere nearby and we, 
perhaps, just do not know an appropriate mathematics to describe it.  

The rest of the paper is organized as follows. In section 2 we, for the purpose
of closeness of this paper, remind construction of non-supersymmetric 
self-dual perturbiner \cite{RS1}.\footnote{A solution, similar to our 
self-dual perturbiner, has been discussed in \cite{Korepin}}
 A key point is a sort of Riemann-Hilbert
problem Eq.(\ref{25}), which is solved upon introducing the so-called ``color
ordering'' (see Eq(\ref{26}) and explanations about it). Interestingly,
the same solution Eq.({\ref{27}) of the same problem Eq.({\ref{25}) was
shown  \cite{RS4} to generate tree form-factors in sin(h)-Gordon theory.
In section 3 we remind $N=3$ super Yang-Mill equations and construct 
the plane wave solution of the linearized field equations. In section 4
we describe the chiral $N=3$ perturbiner. In section 5 the generic (non-chiral)
perturbiner is considered.

\section{Non-supersymmetric self-dual perturbiner \cite{RS1}}
We adopt the spinor notation, so that the connection-form has two indices,
$A_{\alpha \ald}$, $\alpha=1,2 \: {\ald}={\dot 1},{\dot 2}$, so has   
the space-time derivative, $\partial_{\alpha \ald}=
\frac{\partial}{\partial x^{\alpha \ald}}$, and the connection itself,
$\nabla_{\alpha \ald}=\partial_{\alpha \ald}+A_{\alpha \ald}$. The curvature
form, 
\beq
\label{3}
F_{\alpha \ald \beta \bed}=[\nabla_{\alpha \ald},\nabla_{\beta \bed}]
\eeq
has four indices and, being antisymmetric with respect to transposition
${\alpha \ald} \leftrightarrow {\beta \bed}$, decomposes as follows
\beq
\label{4}
F_{\alpha \ald \beta \bed}=\varepsilon_{\alpha \beta}
F_{\ald \bed}+\varepsilon_{\ald \bed} F_{\alpha \beta}
\eeq
where $\varepsilon_{\alpha \beta}$, $\varepsilon_{\ald \bed}$ are the 
standard antisymmetric tensors and the fields
$F_{\ald \bed}$, $F_{\alpha \beta}$ are
symmetric with respect to transposition of indices. The first term in the 
r.h.s. of Eq.(\ref{4}) is 
the self-dual part of the curvature, the second - antiself-dual (the metric
in this notation is $g_{\alpha \ald \beta \bed}=\varepsilon_{\alpha \beta}
\varepsilon_{\ald \bed}$). Then, the self-duality equation is
\beq
\label{5}
 F_{\alpha \beta}=0
\eeq
The (anti)self-duality equation is very well known to be sufficient for the 
Yang-Mills equation to be satisfied.

Construction of the perturbiner (see Introduction) starts with picking 
up a solution of the 
linearized (``free'') field equation. Linearized version of the self-duality
equation reads
\beq
\label{6}
\left(\partial_{\alpha \ald}A_{\beta \bed}-\partial_{\beta \bed}A_{\alpha \ald}
\right)|_{{\rm symmetrized\, in}\: \alpha, \beta}=0  
\eeq
For a plane wave solution (see Eq.(\ref{1})),
\beq
\label{plane}
A_{\alpha \ald}=\epsilon_{\alpha \ald} t e^{k_{\beta \bed}x^{\beta \bed}}
\eeq
Eq.(\ref{6}) gives
\beq
\label{7}
\left(k_{\alpha \ald}\epsilon_{\beta \bed}-k_{\beta \bed}\epsilon_{\alpha \ald}
\right)|_{{\rm symmetrized\, in}\: \alpha, \beta}=0
\eeq
which, in turn, gives that 
\begin{eqnarray}
\label{8}
k_{\alpha \ald}=\ae_{\alpha}{\bar \ae}_{\ald}\nonumber\\
\epsilon^{(+)}_{\alpha \ald}=\frac{q_{\alpha}{\bar \ae}_{\ald}}
{(\ae,q)}
\end{eqnarray}
Eqs.(\ref{8}) mean that $k_{\alpha \ald}$ and $\epsilon_{\alpha \ald}$
are both light-like (remind that the metric is  $g_{\alpha \ald \beta \bed}=
\varepsilon_{\alpha \beta}\varepsilon_{\ald \bed}$), moreover, the dotted
spinor in decomposition of $k_{\alpha \ald}$ and $\epsilon_{\alpha \ald}$
is the same. $q_{\alpha}$ is an arbitrary (reference) spinor defined up to
(linearized) gauge equivalence,
\beq
\label{9}
q_{\alpha} \sim q_{\alpha}+const \cdot \ae_{\alpha}
\eeq
The factor $(\ae,q)$ is introduced for normalization. The brackets 
$(p,q)$ here and below are defined as 
\beq
\label{10}
(p,q)=p^{\alpha}q_{\alpha}=\varepsilon_{\alpha \beta}p^{\alpha}q^{\beta}
\eeq
(indices are raised and lowered with the $\varepsilon$-symbols). What 
concerns to the normalization, antiself-dual plane wave  would have
a polarization 
\begin{equation}
\label{11}
\epsilon^{(-)}_{\alpha \ald}=\frac{\ae_{\alpha}{\bar q}_{\ald}}
{({\bar \ae},{\bar q})} 
\end{equation}
and 
\begin{equation}
\label{12}
\epsilon^{(+)}_{\alpha \ald} \epsilon^{(-)\alpha \ald}=-1  
\end{equation}
Polarizations Eqs.(\ref{8}) and (\ref{11}) can be seen to define positive
and negative helicity states, correspondingly. Notice that the spinors
$\ae_{\alpha}$, ${\bar \ae}_{\ald}$ entering $k_{\alpha \ald}$ Eq.(\ref{8})
can be considered as independent since we are anyway looking for a complex 
solution. At the end, in computation of probabilities, the reality condition
\begin{equation}
\label{14}
{\bar \ae}_{\ald}=\sqrt{-1}\ae_{\alpha}^{\ast}
\end{equation}
should be imposed. 

So, the appropriate solution of the linearized field equations reads
\begin{equation}
  \label{15}
A^{(1)}_{\alpha \ald}=\sum_{J=1}^{L}\frac{q^{J}_{\alpha}
{\bar \ae}^{J}_{\ald}}{({\ae}^{J},q^{J})}{\hej}^{J}   
\end{equation}
where, as in Eq.(\ref{harm}), 
\beq
\label{harm1}
{\hej}_{J}=t_{J}{\ej},\; {\ej}=a_{J}e^{k_{J}x}
\eeq
where $x^{\alpha \ald}$ stands for a space-time coordinate, 
$k^{J}_{\alpha \ald}$ stands for a momentum
of the $J$-th particle, $t_{J}$ stands for a generator of the color 
group, $a_{J}$ is a symbol of annihilation/creation operator of the $J$-th
particle (obeying the nilpotency condition Eq.(\ref{nilpo})).

We are now going to use the twistor construction \cite{Ward} to describe
solutions of the (nonlinear) self-duality equation (\ref{5}).
 
Introduce a couple of complex numbers, $\rho^{\alpha}$, $\alpha=1,2$.
$\rho^{\alpha}$ will also be referred below as the auxiliary spinor.
 $\rho^{\alpha}$, $\alpha=1,2$ can be
considered as homogeneous coordinates on a $CP^{1}$ space. Contracting 
undotted indices of the curvature form $F_{\alpha \ald \beta \bed}$ 
Eq.(\ref{3}) with $\rho$'s one automatically picks up antiself-dual part
of it (see Eq.(\ref{4})) (because the self-dual part is antisymmetric in the 
undotted indices). Hence, the self-duality equation is equivalent to a sort
of zero-curvature condition
\begin{equation}
 \label{16}
[\nabla_{ \ald},\nabla_{\bed}]=0\; {\rm at\, any}\; \rho^{\alpha},\alpha=1,2 
\end{equation}
where $\nabla_{ \ald}=\rho^{\alpha}\nabla_{\alpha \ald}$. Thus, if one 
introduces
\begin{equation}
\label{17}
A_{\ald}=\rho^{\alpha}A_{\alpha \ald}, 
\end{equation}
any self-dual connection form can be (locally) represented as
\begin{equation}
  \label{18}
A_{\ald}=g^{-1} \partial_{\ald} g  
\end{equation}
where $g$ is a group valued function of $\rho$ and $x$ and
$\partial_{\ald}=\rho^{\alpha}\partial_{\alpha \ald}$. All the non-triviality 
of 
the self-duality equation is now encoded in the condition that $g$ must 
depend on ${\rho}$ in such a way that $A_{\ald}$ is a polynomial of degree 1
in ${\rho}$, as in Eq.(\ref{17}). If $g$ is $\rho$-independent, it is a pure 
gauge transformation, as it is seen from  Eq.(\ref{18}).

The above condition on $\rho$-dependence of $g$ is equivalent to condition 
that $g$ is a homogeneous meromorphic function of $\rho$ of degree 0 such
that $A_{\ald}$ from Eq.(\ref{18}) is a homogeneous holomorphic function
of $\rho$ of degree 1 (a homogeneous holomorphic function
of $\rho$ of degree 1 is necessary just linear in $\rho$, as in 
Eq(\ref{17}).). Notice, that
nontrivial (not a pure gauge) $g$ necessary has singularities in $\rho$,
since if it is regular homogeneous of degree 0, then it is just 
$\rho$-independent, that is , a pure gauge.

All above is about $\rho$-dependence of $g$. In the case of perturbiner, the
group-valued function $g$, like the connection-form $A_{\alpha \ald}$
(see definition of the perturbiner in the Introduction), must be polynomial
in harmonics  ${\ej}_{J}, J=1, \ldots, L$, Eq.(\ref{harm1}). First order terms 
of this polynomial,
\begin{equation}
 \label{19}
g^{(1)}_{ptb}(\rho, \{ {\ej} \})=\sum_{J=1}^{L}g_{J}(\rho){\hej}_{J}  
\end{equation}
are fixed by those in $A_{\alpha \ald}$, Eq.(\ref{15}). Expanding 
Eq.(\ref{18}) up to first order in the harmonic ${\ej}_{J}$ and using 
Eq.(\ref{15}) one obtains
\begin{equation}
  \label{20}
g_{J}(\rho)=\frac{(\rho,q^{J})}{(\rho,\ae^{J})(\ae^{J},q^{J})}  
\end{equation}
Eqs.(\ref{19}),(\ref{20}) define the first order terms in expansion of
$g$ in powers of the harmonics  ${\ej}_{J}, J=1, \ldots, L$.

Thus, in terms of $g$, our problem is as follows. We must find polynomial
$g_{ptb}$,
\begin{equation}
  \label{21}
g_{ptb}(\rho, \{ {\ej} \})=1+ g^{(1)}_{ptb}(\rho, \{ {\ej} \})+\:
{\rm higher\, order\, terms\, in\, powers\, of} \: {\ej}'{\rm s} 
\end{equation}
which is a rational function of $\rho^{\alpha}$ of degree $0$, such that 
$A_{\ald}$ from Eq.(\ref{18}) is regular.

Notice that $g^{(1)}_{ptb}$ has first order poles at $\rho^{\alpha}=
\ae_{J}^{\alpha}, J=1, \ldots, L$ (due to factors $(\rho,\ae^{J})=
\varepsilon_{\alpha \beta} \rho^{\alpha} \ae^{\beta}$ in 
denominators, see Eq.(\ref{20})). We remind that $\ae_{J}^{\alpha}$ is
a spinors which appears in decomposition of the corresponding four
momentum $k_{\alpha \ald}$, Eq.(\ref{8}). Importantly, a singular part
of complete $g_{ptb}$ (not only of $g^{(1)}_{ptb}$)
at $\rho^{\alpha}=\ae_{J}^{\alpha}$  is necessary
proportional to the harmonic
${\ej}_{J}$ (that is because $g_{ptb}$ at ${\ej}_{J}=0$ does not contain
any information about the $J$-th particle; form-factors including the 
$J$-th particle will not be generated by the perturbiner at ${\ej}_{J}=0$).
Then, taking into account the nilpotency Eq.(\ref{nilpo}), ${\ej}_{J}^{2}=0$, 
one can show that $g_{ptb}$ may have only simple pole at  
 $\rho^{\alpha}=\ae_{J}^{\alpha}$ for $A_{\ald}$ from Eq.(\ref{18}) to be 
regular at this point. Moreover, residue of $g_{ptb}$ at this point
must obey the condition
\begin{equation}
  \label{22}
\left(\partial_{\ald}(res|_{\rho=\ae_{J}}g \cdot g^{-1})
\right)|_{\rho=\ae_{J}}=0,
\end{equation}
and, according to the rules of the game (see definition of the perturbiner
in the Introduction), Eq.(\ref{22}) must be solved in the form of a
polynomial in the harmonics  ${\ej}_{J}, J=1, \ldots, L$. Clearly, the unique
(up to a color Lee algebra valued constant) solution of Eq.(\ref{22}) reads
\begin{equation}
  \label{23}
res|_{\rho=\ae_{J}}g \cdot g^{-1}|_{\rho=\ae_{J}}={\widehat {const_{J}}}
\cdot {\ej}_{J}  
\end{equation}
(recall, that $\partial_{\ald}|_{\rho=\ae_{J}}=\ae_{J}^{\alpha}
\partial_{\alpha \ald}$). ${\widehat {const}}$ entering Eq.(\ref{23}) is
constant in the sence that it is ${\ej}$-independent. It can be found by
putting all the harmonics ${\ej}$ but ${\ej}_{J}$ to $0$. Then, using
Eqs.(\ref{19}),(\ref{20}), one finds
\begin{equation}
  \label{24}
res|_{\rho=\ae_{J}}g_{J} \cdot t_{J}={\widehat {const_{J}}} 
\end{equation}
with ``one-particle'' $g_{J}$ from Eq.(\ref{20}). 

Eqs.(\ref{23}),(\ref{24}) are seen to be equivalent to the condition that
\begin{equation}
  \label{25}
(1-g_{J}{\hej}_{J})g_{ptb}\; {\rm is\, regular\, at} \rho^{\alpha}=
\ae_{J}^{\alpha}, J=1, \ldots, L 
\end{equation}
(with $g_{J}$ from Eq.(\ref{20})). The condition, expressed by  Eq.(\ref{20}),
defines $g_{ptb}$ uniquely modulo gauge transtformations (that is, modulo 
multiplication by a $\rho$-independent matrix from the right).

To conveniently represent the solution of  Eq.(\ref{25}) let us assume for a 
moment that the generators $t_{J}, J=1, \ldots, L$ defining color states
of the gluons (see Eq.(\ref{15})) belong to a free associative algebra
(that is, there is no relation between them but the associativity 
relation $(t_{J_{1}}t_{J_{2}})t_{J_{3}}=t_{J_{1}}(t_{J_{2}}t_{J_{3}})$).
This means that all monomials of the type of
${\hej}_{J_{1}}{\hej}_{J_{2}} \ldots {\hej}_{J_{L}}$ (${\hej}_{J}$ as in
Eq.(\ref{15})) are linearly independent. $g_{ptb}$ is then uniquely
represented as
\begin{equation}
  \label{26}
g_{ptb}(\rho, \{ {\ej} \})=1+\sum_{J}g_{J}(\rho){\hej}_{J}+
\sum_{J_{1},J_{2}}g_{J_{1},J_{2}}(\rho){\hej}_{J_{1}}{\hej}_{J_{2}}+ \ldots
\end{equation}
Eq.(\ref{25})) is then easily solved for the coefficients
$g_{J_{1},J_{2}, \ldots, J_{L}}$,
\begin{eqnarray}
  \label{27}
g_{J_{1},J_{2}, \ldots, J_{L}}(\rho)=
\frac{(\rho,q^{J_{1}})}{(\rho,\ae^{J_{1}})}
\frac{(\ae^{J_{1}},q^{J_{2}})(\ae^{J_{2}},q^{J_{3}}) \ldots
(\ae^{J_{L-1}},q^{J_{L}})}
{(\ae^{J_{1}},\ae^{J_{2}})(\ae^{J_{2}},\ae^{J_{3}}) \ldots
(\ae^{J_{L-1}},\ae^{J_{L}})}\nonumber\\
\frac{1} {(\ae^{J_{1}},q^{J_{1}})(\ae^{J_{2}},q^{J_{2}}) \ldots
(\ae^{J_{L}},q^{J_{L}})} 
\end{eqnarray}
Eqs.(\ref{26}),(\ref{27}) is a solution of the problem Eq.(\ref{25}).
Of course, it remains to be a  solution if ones introduces back the  
relations between the color group generators $t_{J}, J=1, \ldots, L$.
Any other solution is obtained from Eqs.(\ref{26}),(\ref{27})
multiplicating it by a $\rho$-independent matrix from the right.
\footnote{There is a minor subtlety at this point. When one specifies
 $t_{J}, J=1, \ldots, L$ in Eqs.(\ref{26}),(\ref{27}) to be matrixes
belonging to a gauge Lie algebra, $g_{ptb}$ defined by 
Eqs.(\ref{26}),(\ref{27}) will not necessary belong to the corresponding 
gauge group, only to $GL(\ast)$ instead. It will however be gauge equivalent
(over $GL(\ast)$) to a matrix from the gauge group.}

The connection-form, $A^{ptb}_{\alpha \ald}$, is obtained from
$g^{ptb}$, Eqs.(\ref{26}),(\ref{27}), via Eqs.(\ref{17}),(\ref{18}).
One can do it by a straightforward computation. One can also simplify
 the computation noticing that by construction 
$A^{ptb}_{\ald}=g_{ptb}^{-1} \partial_{\ald} g_{ptb}$ is linear in 
$\rho^{\alpha},\; \alpha=1,2$. Hence  $A^{ptb}_{\alpha \ald}$ can
be found as $\rho$-derivative of $A^{ptb}_{\ald}$ taken at any value of
$\rho$. Choosing all $q$'s in Eq.(\ref{27}) equal each other and equal to a
spinor $q$ (recall that $q$'s are defined up to the gauge freedom
Eq.(\ref{9})) we find  $A^{ptb}_{\alpha \ald}$  
as $\rho$-derivative of $A^{ptb}_{\ald}$ 
at $\rho^{\alpha}=q^{\alpha}$. Since 
 $g^{ptb}|_{\rho^{\alpha}=q^{\alpha}}=1$ (see Eq.(\ref{27})), the computation 
becomes really easy, and one finds
\begin{eqnarray}
\label{28}
A^{ptb}_{\alpha \ald}=\sum_{J=1}^{L}A^{J}_{\alpha \ald}{\hej}_{J}+
\sum_{J_{1}J_{2}}A^{J_{1}J_{2}}_{\alpha \ald}{\hej}_{J_{1}J_{2}}+ \ldots
\nonumber\\
A^{J_{1} \ldots J_{M}}_{\alpha \ald}=
-\frac{q_{\alpha}q^{\beta}k^{J_{1} \ldots J_{M}}_{\beta \ald}}
{(\ae_{J_{1}},q)(\ae_{J_{M}},q)}
\frac{1}{(\ae^{J_{1}},\ae^{J_{2}})(\ae^{J_{2}},\ae^{J_{3}}) \ldots
(\ae^{J_{M-1}},\ae^{J_{M}})}  
\end{eqnarray}
where $ k^{J_{1}J_{2} \ldots J_{L}}_{\alpha \ald}= k^{J_{1}}_{\alpha \ald}+
 k^{J_{2}}_{\alpha \ald}+ \ldots + 
 k^{J_{L}}_{\alpha \ald}$.
One can see that the above choice of $q$'s corresponds to the Lorentz
gauge. Thus $A^{ptb}$ from Eqs.(\ref{28}) is a generating
function (in the sense of Eq.(\ref{2}))
for tree form-factors, or {currents} introduced in
\cite{BerendsGiele}, in the Lorence gauge.

Using $g^{ptb}$ Eqs.(\ref{27}) one easily obtains the prominent Parke-Taylor
amplitudes \cite{ParkeTaylor}. This is done in \cite{RS1} and we shall not 
repeat it here.

\section{N=3 supersymmetric perturbiner; preliminaries}
$N=3$ and $N=4$ Yang-Mills theories are equivalent but N=3 formalism is more
naturally combined with the twistors \cite{Witten} that is why we adopt 
$N=3$ notation. $N=3$ super-space is parametrized by commuting coordinates
$x^{\alpha \ald}$ and by the anticommuting ones $\theta^{\alpha j}$,
${\bth}^{\ald}_{j}$. $\alpha=1,2; \ald={\dot 1},{\dot 2}$ are the Lorentz 
indices, they can be lowered and raised with the antisymmetric 
$\varepsilon$-tensors, $j=1,2,3$ is an {isotopic} index.
Supercharges act in the super-space as
\begin{eqnarray}
  \label{3.0}
Q_{\alpha j}=\frac{\partial}{{\th}^{\alpha j}}-
\frac{1}{2} {\bth}^{\ald}_{j}{\partial}_{\alpha \ald}\nonumber\\
{\bar Q}_{\ald}^{j}=\frac{\partial}{{\bth}^{\ald}_{j}}-
\frac{1}{2} {\th}^{\alpha j}{\partial}_{\alpha \ald}  
\end{eqnarray}
Introduce super-covariant derivatives,
\begin{eqnarray}
  \label{3.1}
D_{\alpha j}=\frac{\partial}{{\th}^{\alpha j}}+
\frac{1}{2} {\bth}^{\ald}_{j}{\partial}_{\alpha \ald}\nonumber\\
{\bar D}_{\ald}^{j}=\frac{\partial}{{\bth}^{\ald}_{j}}+
\frac{1}{2} {\th}^{\alpha j}{\partial}_{\alpha \ald}    
\end{eqnarray}
Introduce also super-connections
\begin{eqnarray}
  \label{3.3}
\nabla_{\alpha j}=D_{\alpha j}+A_{\alpha j}\nonumber\\
{\bar \nabla}_{\ald}^{j}={\bar D}_{\ald}^{j}+{\bar A}_{\ald}^{j}
\end{eqnarray}
where $A_{\alpha j}$ and ${\bar A}_{\ald}^{j}$ are superfields. 

N=3 supersymmetric Yang-Mills equations can be represented as \cite{Witten},
\cite{Sohnius}-\cite{Harnad2}
\begin{eqnarray}
  \label{3.4}
\left(\nabla_{\alpha j}\nabla_{\beta l}+ \nabla_{\beta l}\nabla_{\alpha j}
\right)|_{{\rm symmetrized\, in}\: \alpha, \beta}=0\nonumber\\
\left({\bar \nabla}_{\ald}^{j}{\bar \nabla}_{\bed}^{l}+
{\bar \nabla}_{\bed}^{l}{\bar \nabla}_{\ald}^{j} \right)
|_{{\rm symmetrized\, in}\: \ald, \bed}=0\nonumber\\
\nabla_{\alpha j}{\bar \nabla}_{\bed}^{l}+
{\bar \nabla}_{\bed}^{l}\nabla_{\alpha j}=\delta_{j}^{l}\nabla_{\alpha \bed} 
\end{eqnarray}
(only traceless part of the last equation is, in fact, equation on
 $A_{\alpha j}$ and ${\bar A}_{\ald}^{j}$, the rest is definition of
connection $\nabla_{\alpha \ald}={\partial}_{\alpha \ald}+
A_{\alpha \ald}$). 

Linearization of Eqs.(\ref{3.4}) reads
\begin{eqnarray}
  \label{3.5}
\left( D_{\alpha j}A_{\beta l}+D_{\beta l}A_{\alpha j} \right)
|_{{\rm symmetrized\, in}\: \alpha, \beta}=0\nonumber\\
\left(
{\bar D}_{\ald}^{j}{\bar A}_{\bed}^{l}+{\bar D}_{\bed}^{l}{\bar A}_{\ald}^{j}
\right)|_{{\rm symmetrized\, in}\: \ald, \bed}=0\nonumber\\
D_{\alpha j}{\bar A}_{\bed}^{l}+{\bar D}_{\bed}^{l}A_{\alpha j}=
\delta_{j}^{l}A_{\alpha \bed}  
\end{eqnarray}
As discussed in the Introduction, $N=4$ multiplet splits into two
$N=3$ multiplets, see tables (\ref{N=4}),(\ref{N=3a}),(\ref{N=3b}).

We first describe plane waves corresponding to the highest helicity
states in each multiplet, that is the positive helicity gluon and
$+\frac{1}{2}$ singlet gluino. These plane waves are
\begin{equation}
  \label{3.6}
A_{\alpha j}=\frac{q_{\alpha}({\bth}_{j},{\bar \ae})t}{(\ae,q)}
e^{k_{\alpha \ald}y^{\alpha \ald}},\;{\bar A}_{\ald}^{j}=0  
\end{equation}
(positive helicity gluon)
\begin{equation}
  \label{3.7}
A_{\alpha j}=0,\; {\bar A}_{\ald}^{j}=\frac{{\bar q_{\ald}} \frac{1}{2} 
{\varepsilon}^{jlm}
({\bth}_{l},{\bar \ae})({\bth}_{m},{\bar \ae})t}{({\bar \ae},{\bar q})}
e^{k_{\alpha \ald}y^{\alpha \ald}}  
\end{equation}
($+\frac{1}{2}$ singlet gluino).
In these equations $y^{\alpha \ald}=x^{\alpha \ald}+
\frac{1}{2}{\th}^{\alpha j}{\bth}^{\ald}_{j}$ is a {\it chiral} coordinate,
defined so that 
\begin{equation}
  \label{3.8}
{\bar D}_{\bed}^{l}y^{\alpha \ald}=0,  
\end{equation}
the bracket $(a,b)$, as before, stands for contraction of $a$ and $b$
with the ${\varepsilon}$-tensor, Eq.(\ref{10}), ${\varepsilon}^{jlm}$ is
the totally antisymmetric tensor in the isotopic space,
$k^{\alpha \ald}=\ae^{\alpha} {\bar \ae}^{\ald}$ is a (light-like)
four-momentum, $t$ is a gauge Lie algebra generator,
$q_{\alpha}$ and ${\bar q}_{\ald}$ are the reference spinors,
they are defined modulo gauge equivalence
\begin{eqnarray}
  \label{3.9}
q_{\alpha} \sim q_{\alpha}+ const \cdot \ae_{\alpha}\nonumber\\
{\bar q}_{\ald} \sim {\bar q}_{\ald}+ const \cdot {\bar \ae}_{\ald}.  
\end{eqnarray}
One can check by a straightforward computation that the plane waves,
Eqs.(\ref{3.6}), (\ref{3.7}) go through the linearized field equations
Eqs.(\ref{3.5}). One can also check that 
\begin{eqnarray}
  \label{3.10}
Q_{\alpha j}A^{\beta l}=0\nonumber\\
Q_{\alpha j}{\bar A}_{\bed}^{l}=0  
\end{eqnarray}
and
\begin{eqnarray}
  \label{3.11}
({\bar Q}^{j},{\bar \ae})A^{\beta l}=0\nonumber\\
({\bar Q}^{j},{\bar \ae}){\bar A}_{\bed}^{l}=0  
\end{eqnarray}
on the both states Eqs.(\ref{3.6}) and Eqs.(\ref{3.7}).
 Eqs.(\ref{3.10}) means that these states are highest states in the 
multiplets, while  Eqs.(\ref{3.11}) mean that half of the supercharges act
trivially on the whole multiplets (since these states are massless).

Acting with the ${\bar Q}$ charges on the states 
 Eqs.(\ref{3.6}) and Eqs.(\ref{3.7}) one can obtain plane waves 
corresponding to all states in the tables
 Eqs.(\ref{N=3a}) and Eqs.(\ref{N=3b}). We shall, however do a bit differently.
Instead of plane waves of the type of 
$e^{k_{\alpha \ald}y^{\alpha \ald}}$ we shall use plane waves of the type of
\begin{equation}
  \label{3.12}
e^{k_{\alpha \ald}y^{\alpha \ald}+({\th}^{j},{\ae}) \chi_{j}}  
\end{equation}
which are proper states for the supercharges $Q_{\alpha j}$,
\begin{equation}
  \label{3.14}
Q_{\beta l}e^{k_{\alpha \ald}y^{\alpha \ald}+({\th}^{j},{\ae}) \chi_{j}}
={\ae}_{\beta}{\chi}_{l}
e^{k_{\alpha \ald}y^{\alpha \ald}+({\th}^{j},{\ae}) \chi_{j}}  
\end{equation}
where ${\chi}_{l}, l=1,2,3$ are Grassmann variables; they are super-partners
of momentum $k_{\alpha \ald}$, so they will be referred to as 
{\it {momentino}}.

Now the multiplets  Eqs.(\ref{N=3a}) and Eqs.(\ref{N=3b}) can be organized
as 
\begin{equation}
\label{3.15}
A_{{\alpha} j}=
\frac{q_{\alpha}(({\bth}_{j},{\bar \ae})+\chi_{j})t}{(\ae,q)}
e^{k_{\alpha \ald}y^{\alpha \ald}+
({\th}^{j},{\ae}) \chi_{j}}, \; {\bar A}_{\ald}^{j}=0  
\end{equation}
(positive helicity gluon)
\begin{equation}
  \label{3.16}
A_{\alpha j}=0, \;
{\bar A}_{\ald}^{j}=\frac{{\bar q} \frac{1}{2} {\varepsilon}^{jlm}
(({\bth}_{l},{\bar \ae})+\chi_{l})
(({\bth}_{m},{\bar \ae})+\chi_{m})t}{({\bar \ae},{\bar q})}
e^{k_{\alpha \ald}y^{\alpha \ald}+
({\th}^{j},{\ae}) \chi_{j}}  
\end{equation}
Various members of the multiplets  Eqs.(\ref{N=3a}) and Eqs.(\ref{N=3b})
arise as coefficients in the Taylor expansion of 
 Eqs.(\ref{3.15}) and Eqs.(\ref{3.16}) in powers of ${\chi}_{j}$. For
example, the plane wave solution corresponding to the 
negative helicity gluon state arises as coefficient at
$\frac{1}{3!} {\varepsilon}^{jlm}\chi_{j}\chi_{l}\chi_{m}$ in the expansion 
of Eq.(\ref{3.16}).

\section{Chiral N=3 supersymmetric perturbiner}
In this section we construct perturbiner which generates only form-factors
with all asymptotic states belonging to the same type of the $N=3$ 
super-multiplets,
say, the one from the table (\ref{N=3a}) (no kinematical restrictions are 
assumed, that is, the set of momenta is arbitrary).
 According to the definition of the perturbiner
(see Introduction),
one first picks up a solution of the linearized field equations 
(\ref{3.5}) in the form of a superposition of plane waves of the 
type of Eq.(\ref{3.15}),
\begin{equation}
  \label{4.1}
A^{(1)}_{\alpha j}=\sum_{J=1}^{L}
\frac{q^{J}_{\alpha}(({\bth}_{j},{\bar \ae^{J}})+\chi^{J}_{j})t}
{(\ae^{J},q^{J})}
{\shej}_{J},\; {\bar A}_{\ald}^{j}=0    
\end{equation}
where 
\begin{equation}
  \label{sharm}
{\shej}_{J}=t_{J}{\sej}, \;
{\sej}=a_{J}e^{k_{\alpha \ald}y^{\alpha \ald}+
({\th}^{j},{\ae}^{J}) \chi^{J}_{j}},  
\end{equation}
$a^{J}$ is a commuting nilpotent ( see Eq.(\ref{nilpo})) symbol of 
annihilation/creation operator of the $J$-th particle, 
${\chi}^{J}_{j}$ stands for the momentino of the $J$-th particle
(see Eq(\ref{3.14})),
other notations
are as in Eqs.(\ref{3.6}),(\ref{3.7}).

Then one looks for a solution of the field equations (\ref{3.4}),
which is polynomial in the super-harmonics 
${\sej}_{J}, J=1, \ldots, L$ Eq.(\ref{sharm}), and whose first order term is 
as in Eq.(\ref{4.1}). To describe this solution, introduce again an 
auxiliary $CP^{1}$ space with homogeneous coordinates 
$\rho^{\alpha}, \alpha=1,2$. Introduce, also, $D_{j}$, $\nabla_{j}$
and $A_{j}$ as
\begin{eqnarray}
  \label{4.2}
D_{j}=\rho^{\alpha}D_{\alpha j}\nonumber\\
\nabla_{j}=\rho^{\alpha}\nabla_{\alpha j}\nonumber\\
A_{j}=\rho^{\alpha}A_{\alpha j},   
\end{eqnarray}
Then, as in the self-dual case (section 2), first equation of 
Eqs.(\ref{3.4}) is represented as a zero-curvature condition
by contracting its Lorence indices with $\rho$'s. Hence, the first
equation is (locally) solved as
\begin{equation}
  \label{4.3}
A_{j}=g^{-1} D_{j} g  
\end{equation}
where (superfield) $g$ on r.h.s. is (similarly to Eq.(\ref{18})) 
a group valued rational homogeneous function of 
$\rho^{\alpha}, \alpha=1,2$ of degree $0$, such that (superfield)
$A_{j}$ is a regular homogeneous function of 
$\rho^{\alpha}, \alpha=1,2$ of degree $0$ of degree $1$. The rest
equations of Eqs.(\ref{3.4}) are solved provided
\begin{equation}
  \label{4.4}
{\bar D}_{\ald}^{j}g=0
\end{equation}
According to the rules of the game, $g^{sptb}$ is sought for
as a polynomial in the super-harmonics ${\sej}_{J}, J=1, \ldots, L$
Eq.(\ref{sharm}), first order term of the polynomial being defined
by the one in $A_{\alpha j}$ Eq.(\ref{4.1}) via Eq.(\ref{4.3}).
All steps in constructing such $g^{sptb}$ are parallel to the ones
in section 2. Moreover, amusingly, the resulting $g^{sptb}$ is
given by the same formulae as  $g^{ptb}$ Eq.(\ref{26}),(\ref{27})
with the substitution 
\begin{equation}
  \label{4.5}
{\hej} {\rightarrow} {\shej}
\end{equation}
(Eq.(\ref{4.4}) is satisfied because 
${\bar D}^{j} {\shej}=0$)

Clearly, $g^{sptb}$ has the same singularities in the auxiliary
$CP^{1}$ space as the non-supersymmetric self-dual perturbiner
$g^{ptb}$, Eq.(\ref{26}),(\ref{27}), namely, it has simple poles
at   $\rho^{\alpha}=\ae_{J}^{\alpha}, J=1, \ldots, L$ where
$\ae_{J}^{\alpha}$ is the spinor appearing in decomposition
of momentum $k^{J}_{\alpha \ald}$ of the $J$-th particle,
(see Eq.(\ref{8})).

Finally, the generating functions for form-factors of the superfields
$A_{\alpha j},\;{\bar A}_{\ald}^{j}$  are obtained from Eqs.(\ref{4.3})
(computation is parallel to the one in section 2),
\begin{eqnarray}
\label{4.6}
A^{sptb}_{\alpha j}=\sum_{J=1}^{L}A^{J}_{\alpha j}{\hej}_{J}+
\sum{J_{1}J_{2}}A^{J_{1}J_{2}}_{\alpha j}{\hej}_{J_{1}J_{2}}+ \ldots;
\: {\bar A}^{sptb \: j}_{\ald}=0
\nonumber\\
A^{J_{1} \ldots J_{M}}_{\alpha j}=-\frac{q_{\alpha}
q^{\beta} \left([\ae_{\beta} \chi_{j}]^{J_{1} \ldots J_{M}}+
{\bth}^{\ald}k^{J_{1} \ldots J_{M}}_{\beta \ald} \right)}
{(\ae_{J_{1}},q)(\ae_{J_{M}},q)} \cdot \nonumber\\
\frac{1}{(\ae^{J_{1}},\ae^{J_{2}})(\ae^{J_{2}},\ae^{J_{3}}) \ldots
(\ae^{J_{M-1}},\ae^{J_{M}})}   
\end{eqnarray}
where $[\ae_{\beta} \chi_{j}]^{J_{1} \ldots J_{M}}=
\ae_{\beta}^{J_{1}}\chi_{j}^{J_{1}}+ \ldots +
\ae_{\beta}^{J_{M}} \chi_{j}^{J_{M}} $.

\section{Generic (nonchiral) N=3 perturbiner}
Nonchiral $N=3$ supersymmetric perturbiner is a generating function for 
all tree
form-factors in the $N=3$ ($N=4$) supersymmetric Yang-Mills theory. 
According to the rules
of the game, the solution of the linearized field equations must
include now both types of harmonics, Eqs.(\ref{3.15}),(\ref{3.16}),
\begin{eqnarray}
  \label{5.1}
A^{(1)}_{\alpha j}=\sum_{J=1}^{L}
\frac{q^{J}_{\alpha}(({\bth}_{j},{\bar \ae^{J}})+\chi^{J}_{j})t}
{(\ae^{J},q^{J})}
{\shej}_{J}\nonumber\\
{\bar A}^{(1) j}_{\ald}=\sum_{J=1}^{L}b_{J}\frac{{\bar q}_{\ald}^{J} 
\frac{1}{2} {\varepsilon}^{jlm}
(({\bth}_{l},{\bar \ae}^{J})+\chi^{J}_{l})
(({\bth}_{m},{\bar \ae}^{J})+\chi^{J}_{m})t}{({\bar \ae}^{J},{\bar q}^{J})}    
{\shej}_{J}
\end{eqnarray}
where $b_{J}$ is an anticommuting nilpotent symbol, all other notations 
are the same as in Eq.(\ref{4.1}). 

The nonchiral perturbiner is a solution of the field equations 
Eqs.(\ref{3.4}), polynomial in the harmonics ${\sej}_{J}, J=1, \ldots, L$,
Eq.(\ref{sharm}), first order term of the polynomial being just
$A^{(1)}$ as in Eq.(\ref{4.1}). We again use a  twistor
construction \cite{Witten} to describe
solutions of Eqs.(\ref{3.4}). To this end, introduce two couples of 
complex numbers, ${\rho}^{\alpha}, \alpha=1,2$, ${\bar \rho}^{\ald}, 
\ald={\bar 1}, {\bar 2}$ which can be viewed on as homogeneous coordinates
on $CP^{1} \times CP^{1}$ space. Contracting all Lorence indices in 
Eqs.(\ref{3.4}) with these ${\rho}$'s, ${\bar \rho}$'s one 
again obtains a sort of zero-curvature condition
\begin{eqnarray}
  \label{5.2}
\nabla_{j}\nabla_{l}+\nabla_{l}\nabla_{j}=0\nonumber\\
{\bar \nabla}^{j}{\bar \nabla}^{l}+
{\bar \nabla}^{l}{\bar \nabla}^{j}=0\nonumber\\
\nabla_{j}{\bar \nabla}^{l}+{\bar \nabla}^{l}\nabla_{j}=
\delta_{j}^{l}\nabla  
\end{eqnarray}
where 
\begin{eqnarray}
  \label{5.3}
\nabla_{j}=\rho^{\alpha}\nabla_{\alpha j}\nonumber\\
{\bar \nabla}^{j}=
{\bar \rho}^{\ald}{\bar \nabla}_{\ald}^{j}\nonumber\\
\nabla=\rho^{\alpha}{\bar \rho}^{\ald}\nabla_{\alpha \ald}  
\end{eqnarray}
Clearly, Eqs.(\ref{3.4}) and Eqs.(\ref{5.2}) are equivalent provided
Eqs.(\ref{5.2}) are solved identically in  ${\rho}$, ${\bar \rho}$.
Eqs.(\ref{5.2}) are locally solved as
\begin{eqnarray}
  \label{5.4}
A_{j}=g^{-1} D_{j} g\nonumber\\
{\bar A}^{j}=g^{-1} {\bar D}^{j} g  
\end{eqnarray}
where 
\begin{eqnarray}
  \label{5.5}
A_{j}=\rho^{\alpha}A_{\alpha j}\nonumber\\
D_{j}=\rho^{\alpha}D_{\alpha j}\nonumber\\
{\bar A}^{j}={\bar \rho}^{\ald}{\bar A}_{\ald}^{j}\nonumber\\
{\bar D}^{j}=
{\bar \rho}^{\ald}{\bar D}_{\ald}^{j}  
\end{eqnarray}
and  ${\rho}$, ${\bar \rho}$ dependence of $g$ must be such that 
$A_{j}$, ${\bar A}^{j}$ are just linear in  ${\rho}$, ${\bar \rho}$,
as in Eqs.(\ref{5.5}). More definitely, (superfield) $g$ is a
meromorphic homogeneous function of  ${\rho}$, ${\bar \rho}$ of 
degree $(0,0)$ such that (superfield) $A_{j}$ Eqs.(\ref{5.5}) 
is a holomorphic homogeneous function of  ${\rho}$, ${\bar \rho}$
of degree $(1,0)$ and  (superfield) ${\bar A}^{j}$ Eqs.(\ref{5.5})
is a holomorphic homogeneous function of  ${\rho}$, ${\bar \rho}$
of degree $(0,1)$, as in Eqs.(\ref{5.5}).

Again, in the case of perturbiner, $g^{ncsptb}$ must be polynomial
in the super-harmonics ${\sej}_{J}, J=1, \ldots, L$,
Eq.(\ref{sharm}). First order terms of this polynomial,
\begin{equation}
 \label{5.6}
g^{(1)}_{ncsptb}(\rho, {\bar \rho}, \{ {\sej} \})=
\sum_{J=1}^{L}g^{ncsptb}_{J}(\rho, {\bar \rho}){\shej}_{J}  
\end{equation} 
are fixed by those in $A$'s  Eqs.(\ref{5.1}) via 
Eqs.(\ref{5.4}) (analogously to 
 Eqs.(\ref{19}), Eqs.(\ref{20}).
 Expanding 
Eq.(\ref{5.4}) up to first order in the harmonic ${\sej}_{J}$ and using 
Eq.(\ref{5.1}),Eq.(\ref{5.6}) one obtains
\begin{eqnarray}
  \label{5.7}
g^{ncsptb}_{J}(\rho,{\bar \rho} )=
\frac{(\rho,q^{J})}{(\rho,\ae^{J})(\ae^{J},q^{J})}+
b^{J}\frac{({\bar \rho},{\bar q}^{J})}{({\bar \rho},{\bar \ae}^{J})
({\bar \ae}^{J},{\bar q}^{J})} \cdot\nonumber\\  
\frac{1}{3!}{\varepsilon}^{jlm}
(({\bth}_{l},{\bar \ae}^{J})+\chi^{J}_{j})
(({\bth}_{l},{\bar \ae}^{J})+\chi^{J}_{l})
(({\bth}_{m},{\bar \ae}^{J})+\chi^{J}_{m})
\end{eqnarray}
Thus, according to Eqs.(\ref{5.6}), Eq.(\ref{5.7}), the first order
terms in $g^{ncsptb}$ have
simple poles at surfaces
\begin{equation}
  \label{5.8}
\rho_{\alpha}={\ae}^{J}_{\alpha},\; J=1, \ldots, L  
\end{equation}
and at surfaces
\begin{equation}
  \label{5.9}
{\bar \rho}_{\ald}={\bar \ae}^{J}_{\ald},\; J=1, \ldots, L  
\end{equation}
in the $CP^{1} \times CP^{1}$ space parametrized by  ${\rho}$, ${\bar \rho}$.
An analysis shows that for $A,\;{\bar A}$ from Eqs.(\ref{5.4}) to be regular,
the higher order terms in the polynomial $g_{ncsptb}$ may have only simple 
poles at the surfaces  Eqs.(\ref{5.8}),(\ref{5.9}) and also at the
surfaces $C_{\cal M}$ in  $CP^{1} \times CP^{1}$
\begin{equation}
  \label{5.10}
C_{\cal M}:\{
\rho^{\alpha} {\bar \rho}^{\ald} k^{\cal M}_{\alpha \ald}=0 \}
\end{equation}
where $ k^{\cal M}_{\alpha \ald}=\sum_{J \in {\cal M}} 
k^{J}_{\alpha \ald}$, and ${\cal M}$ is a subset of the set $J=1, \ldots, L$.
That is, any linear combination of momenta of the asymptotic states
included in the perturbiner defines a surface in $CP^{1} \times CP^{1}$
at which  $g_{ncsptb}$ has a simple pole. Notice that due to the 
non-resonantness condition (see Introduction) the surfaces  $C_{\cal M}$
 Eq.(\ref{5.10}) with ${\cal M}$ including more than one element never
reduces to the ones of the type of  Eqs.(\ref{5.8}),(\ref{5.9}).
Furthermore, the regularity of $A,\;{\bar A}$ from Eqs.(\ref{5.4})
dictates a condition on residues of  $g_{ncsptb}$ at the
surfaces $C_{\cal M}$, namely
\begin{eqnarray}
  \label{5.11}
\left( D_{j}(res|_{C_{\cal M}}g_{ncsptb} \cdot 
g^{-1}_{ncsptb})\right)|_{C_{\cal M}}=0\nonumber\\
\left({\bar D}^{j}
res|_{C_{\cal M}}g_{ncsptb} \cdot g^{-1}_{ncsptb})\right)|_{C_{\cal M}}=0
\end{eqnarray}
where $D_{j}$, ${\bar D}^{j}$ are as in  Eqs.(\ref{5.5}) and notation
$|_{C_{\cal M}}$ in  Eqs.(\ref{5.11}) indicates restriction
at the surface  $C_{\cal M}$  Eq.(\ref{5.10}). It is again convenient to 
use the expansion of  $g_{ncsptb}$ in the color ordered monomials
(see  Eq.(\ref{26}) and explanations about it),
\begin{equation}
  \label{5.12}
g_{ncsptb}(\rho, {\bar \rho}, \{ {\sej} \})=1+
\sum_{J}g^{ncsptb}_{J}(\rho {\bar \rho}){\shej}_{J}+
\sum_{J_{1},J_{2}}g^{ncsptb}_{J_{1},J_{2}}(\rho, {\bar \rho})
{\shej}_{J_{1}}{\shej}_{J{2}}+ \ldots
\end{equation}
The singularity structure of  $g_{ncsptb}$  Eqs.(\ref{5.10}),(\ref{5.11})
dictates the following form of coefficients 
$g^{ncsptb}_{J_{1},J_{2}, \ldots, J_{L}}(\rho,{\bar \rho})$,
\begin{equation}
  \label{5.14}
g^{ncsptb}_{J_{1},J_{2}, \ldots, J_{M}}(\rho,{\bar \rho})=
\frac{P^{J_{1}J_{2} \ldots J_{M}}(\rho,{\bar \rho},{\bth})}
{(\rho^{\alpha} {\bar \rho}^{\ald} k^{J_{1}}_{\alpha \ald})
(\rho^{\alpha} {\bar \rho}^{\ald} k^{J_{1}J_{2}}_{\alpha \ald})
\ldots
(\rho^{\alpha} {\bar \rho}^{\ald} k^{J_{1}J_{2} 
\ldots J_{M}}_{\alpha \ald})}  
\end{equation}
where, we remind, 
$ k^{J_{1}J_{2} \ldots J_{M}}_{\alpha \ald}= k^{J_{1}}_{\alpha \ald}+
 k^{J_{2}}_{\alpha \ald}+ \ldots + 
 k^{J_{M}}_{\alpha \ald}$,
and $P^{J_{1}J_{2} \ldots J_{M}}(\rho,{\bar \rho},{\bth})$ is a 
polynomial in  ${\rho}$, ${\bar \rho}$ of degree (M,M) (so, that
$g^{ncsptb}_{J_{1},J_{2}, \ldots, J_{M}}(\rho,{\bar \rho})$ is a rational
function of  ${\rho}$, ${\bar \rho}$ of degree $(0,0)$). We have explicitly 
indicated only  ${\rho}$, ${\bar \rho}$ and ${\bth}$ dependence of these
polynomials. They, of course, depend on quantum numbers of $J_{1}$-th,
$J_{2}$-th,...,$J_{M}$-th particles, such as momenta, momentinos
etc. 

Now, the problem of constructing $g_{ncsptb}$, and thus, the problem of 
computing all tree form-factors in (supersymmetric) Yang-Mills theory,
will be formulated as a problem of constructing a set of polynomials
$P^{J_{1}J_{2} \ldots J_{M}}(\rho,{\bar \rho},{\bth})$,
$M=1, \ldots, L$. From  Eqs.(\ref{5.11}),(\ref{5.12}) and (\ref{5.14}),
crucially using the nilpotency, ${\sej}^{2}=0$, and linear independence of 
various color ordered products of ${\sej}$'s, one obtains
\begin{eqnarray}
  \label{5.15i}
\left(\rho^{\alpha}([{\ae}_{\alpha}{\chi}_{j}]^{J_{1} \ldots J_{M}}+
{\bth}^{\ald}_{j} k^{J_{1}J_{2} \ldots J_{M}}_{\alpha \ald})
P^{J_{1}J_{2} \ldots J_{M}}\right)|_{C_{J_{1}J_{2} \ldots J_{M}}} \
=0\nonumber\\  
\left({\bar \rho}^{\ald}\frac{\partial}{{\bth}^{\ald}_{j}}
P^{J_{1}J_{2} \ldots J_{M}}\right)|_{C_{J_{1}J_{2} \ldots J_{M}}}=0
\end{eqnarray}
\begin{equation}
  \label{5.15ii}
P^{J_{1}J_{2} \ldots J_{M}}|_{C_{J_{1}J_{2} \ldots J_{I}}}=
P^{J_{1}J_{2} \ldots J_{I}}|_{C_{J_{1}J_{2} \ldots J_{I}}}
P^{J_{I+1} \ldots J_{M}}|_{C_{J_{1}J_{2} \ldots J_{I}}},\;
I<M,
\end{equation}
and from  Eqs.(\ref{5.7}) one sees that 
\begin{eqnarray}
  \label{5.15iii}
P^{J}=\frac{(\rho,q^{J})({\bar \rho},{\bar q}^{J})}
{(\ae^{J},q^{J})}+
b^{J}\frac{({\bar \rho},{\bar q}^{J})(\rho,\ae^{J})}
{({\bar \ae}^{J},{\bar q}^{J})} \cdot \nonumber\\
\frac{1}{3!}{\varepsilon}^{jlm}
(({\bth}_{j},{\bar \ae}^{J})+\chi^{J}_{j})
(({\bth}_{l},{\bar \ae}^{J})+\chi^{J}_{l})
(({\bth}_{m},{\bar \ae}^{J})+\chi^{J}_{m})  
\end{eqnarray}
We remind that 
$P^{J_{1}J_{2} \ldots J_{M}}(\rho,{\bar \rho},{\bth})$, 
$M=1, \ldots, L$ are homogeneous polynomials on 
 $CP^{1} \times CP^{1}$ space, parametrized by  
${\rho}^{\alpha}, \alpha=1,2$, ${\bar \rho}^{\ald}, \ald={\dot 1},{\dot 2}$,
of degree (M,M).
$ k^{J_{1}J_{2} \ldots J_{L}}_{\alpha \ald}= k^{J_{1}}_{\alpha \ald}+
 k^{J_{2}}_{\alpha \ald}+ \ldots + 
 k^{J_{L}}_{\alpha \ald}$, ($k^{J}_{\alpha \ald}=\ae^{J}_{\alpha}
{\bar \ae}^{J}_{\ald}$ is momentum of the $J$-th particle),
$[\ae_{\beta} \chi_{j}]^{J_{1} \ldots J_{M}}=
\ae_{\beta}^{J_{1}} \chi_{j}^{J_{1}}+
\ae_{\beta}^{J_{2}}\chi_{j}^{J_{2}}+ \ldots +
\ae_{\beta}^{J_{M}}\chi_{j}^{J_{M}}$.   
Notation $P|_{C}$ means that a polynomial $P$ is
restricted on a surface $C$. $C_{J_{1}J_{2} \ldots J_{M}}$ are surfaces 
in  $CP^{1} \times CP^{1}$ defined by  Eqs.(\ref{5.10}). Due to 
the non-resonantness condition (that a sum of the type of
$ k^{J_{1}}_{\alpha \ald}+
 k^{J_{2}}_{\alpha \ald}+ \ldots + 
 k^{J_{L}}_{\alpha \ald}$ is never light-like when there is more than one
term) only surfaces $C_{J}$ are reducible, as in Eqs.(\ref{5.8}),(\ref{5.9}).
Notice that the surfaces 
$C_{J_{1}J_{2} \ldots J_{I}}$ and $C_{J_{I+1}J_{2} \ldots J_{L}}$,
$I<L$ intersect at two points in  $CP^{1} \times CP^{1}$, defined by the system
\begin{eqnarray}
\label{points}
\rho^{\alpha} {\bar \rho}^{\ald} k^{J_{1}J_{2} \ldots J_{I}}_{\alpha \ald}=0,
\nonumber\\
\rho^{\alpha} {\bar \rho}^{\ald} k^{J_{I+1}J_{2} \ldots J_{L}}_{\alpha \ald}
=0  
\end{eqnarray}
and at every one of these intersection points three surfaces meet -
$C_{J_{1}J_{2} \ldots J_{I}}$, $C_{J_{I+1}J_{2} \ldots J_{L}}$ and
$C_{J_{1}J_{2} \ldots J_{L}}$.
The bracket of the type of (a,b) has been defined in Eq.(\ref{10}).
$\theta^{\alpha j}$, ${\bth}^{\ald}_{j}$, 
$\alpha=1,2; \ald={\dot 1},{\dot 2}$, $j=1,2,3$ are Grassmann variables.
$\chi^{J}_{j}$, momentino of the $J$-th particle, has been introduced
in  Eqs.(\ref{3.12}),(\ref{3.14}).
$q_{\alpha}$ and ${\bar q}_{\ald}$ are the reference spinors,
they are defined modulo gauge equivalence
$q_{\alpha} \sim q_{\alpha}+const \cdot \ae_{\alpha}, \;
{\bar q}_{\ald} \sim {\bar q}_{\ald}+const \cdot{\bar \ae}_{\ald}$
(one may use this freedom in solving the problem  Eqs.(\ref{5.15i}),
(\ref{5.15ii}),(\ref{5.15iii})). 

We shall now explain how this problem,  Eqs.(\ref{5.15i}),
(\ref{5.15ii}),(\ref{5.15iii}), can be solved iteratively. Namely,
we explain how to find polynomial $P^{J_{1}J_{2} \ldots J_{L}}$,
provided all polynomials  $P^{J_{1}J_{2} \ldots J_{I}}$, $I<L$
are known (in this sense Eq.(\ref{5.15iii}) is a first step of 
the iteration). 

Consider, first, the polynomials restricted on ``their own''
surfaces, that is, introduce
\begin{equation}
  \label{5.16}
{\tilde P}^{J_{1}J_{2} \ldots J_{L}}=
P^{J_{1}J_{2} \ldots J_{L}}|_{C_{J_{1}J_{2} \ldots J_{L}}}.  
\end{equation}
${\tilde P}^{J_{1}J_{2} \ldots J_{L}}$ is a homogeneous polynomial
on $C_{J_{1}J_{2} \ldots J_{L}}$ of degree $2L$, and hence it has
$2L+1$ degrees of freedom.

Equations (\ref{5.15ii}) on the restricted polynomials 
reduce  to 
\begin{eqnarray}
  \label{5.17}
{\tilde P}^{J_{1}J_{2} \ldots J_{L}}|_
{Z_{i}(C_{J_{1}J_{2} \ldots J_{I}} \cap C_{J_{I+1} \ldots J_{L}})}=
P^{J_{1}J_{2} \ldots 
J_{I}}|_{Z_{i}(C_{J_{1}J_{2} \ldots J_{I}} \cap C_{J_{I+1} \ldots J_{L}})}
\nonumber\\
P^{J_{I+1} \ldots 
J_{L}}|_{Z_{i}(C_{J_{1}J_{2} \ldots J_{I}} \cap C_{J_{I+1} \ldots J_{L}})},\;
I<L,
\end{eqnarray}
where $Z_{i}(C_{J_{1}J_{2} \ldots J_{I}} \cap C_{J_{I+1} \ldots J_{L}}), i=1,2$
stand for the intersection points of the surfaces 
$C_{J_{1}J_{2} \ldots J_{I}}$ and  $C_{J_{I+1} \ldots J_{L}}$, 
Eq.(\ref{points}).
Thus Eq.(\ref{17}) defines polynomial ${\tilde P}^{J_{1}J_{2} \ldots J_{L}}$
at these $2(L-1)$ intersection points.

Equations (\ref{5.15i}) tell that the polynomial 
 ${\tilde P}^{J_{1}J_{2} \ldots J_{L}}$ must be of the form
 \begin{eqnarray}
   \label{5.19}
{\tilde P}^{J_{1}J_{2} \ldots J_{L}}(\rho,{\bar \rho},{\bth})=
\frac{1}{3!}{\varepsilon}^{jlm}
\left(\rho^{\alpha}([{\ae}_{\alpha}{\chi}_{j}]^{J_{1} \ldots J_{M}} +
{\bth}^{\ald}_{j} k^{J_{1}J_{2} \ldots J_{M}}_{\alpha \ald}) \right)
\nonumber\\ 
\left(\rho^{\alpha}([{\ae}_{\alpha}{\chi}_{l}]^{J_{1} \ldots J_{M}}+
{\bth}^{\ald}_{l} k^{J_{1}J_{2} \ldots J_{M}}_{\alpha \ald}) \right)
\left(\rho^{\alpha}([{\ae}_{\alpha}{\chi}_{m}]^{J_{1} \ldots J_{M}}+
{\bth}^{\ald}_{m} k^{J_{1}J_{2} \ldots J_{M}}_{\alpha \ald}) \right)
\nonumber\\ 
R^{J_{1}J_{2} \ldots J_{L}}(\rho,{\bar \rho})|_{C_{J_{1}J_{2} \ldots J_{L}}}  
 \end{eqnarray}
where 
$R^{J_{1}J_{2} \ldots J_{L}}(\rho,{\bar \rho})|_{C_{J_{1}J_{2} \ldots J_{L}}}$
is a polynomial on the surface $C_{J_{1}J_{2} \ldots J_{L}}$
of degree $2L-3$. Thus, the number of degrees of freedom
in the polynomial 
$R^{J_{1}J_{2} \ldots J_{I}}(\rho,{\bar \rho})|_{C_{J_{1}J_{2} \ldots J_{L}}}$,
and so in the  polynomial
 ${\tilde P}^{J_{1}J_{2} \ldots J_{I}}$, is $2L-2$, that is, precisely
the number of points at which the polynomial is defined according to 
Eqs.(\ref{5.17})! (Eqs.(\ref{5.19}) and  (\ref{5.17}) are  
compatible due to nilpotency of
$\theta^{\alpha j}$, ${\bth}^{\ald}_{j}$ and $\chi^{J}_{j}$).

Once restriction of the polynomial 
 $P^{J_{1}J_{2} \ldots J_{L}}$ on the surface $C_{J_{1}J_{2} \ldots J_{L}}$
is found and restrictions of the polynomial  $P^{J_{1}J_{2} \ldots J_{L}}$
on the surfaces  $C_{J_{1}J_{2} \ldots J_{M}}$, $M<L$ are known due to
equations (\ref{5.15ii}), the polynomial 
 $P^{J_{1}J_{2} \ldots J_{L}}$ is fixed modulo a polynomial which is zero at 
surfaces  $C_{J_{1}J_{2} \ldots J_{M}}$, $M=1, \ldots ,L$, that is precisely
modulo a polynomial in the denominator of  
$g^{ncsptb}_{J_{1},J_{2}, \ldots, J_{L}}$, Eq.(\ref{5.14}), that is, modulo
the gauge freedom.

The described iterative procedure can, in principle, be used as an
alternative to the usual perturbation theory, and it might even be 
the more economical alternative, but
we shall not try here to prove its efficiency. Instead we would
like to express our hope that Eqs.(\ref{5.15i}),
(\ref{5.15ii}),(\ref{5.15iii}) can be, in some sense, solved completely.
By the way,  Eqs.(\ref{5.15ii}),(\ref{5.15iii})
allow such a complete solution up to a freedom which is to be fixed
by  Eqs.(\ref{5.15i}) (or  Eqs.(\ref{5.19}), namely
\[ P^{J_{1}J_{2} \ldots J_{L}}=\det \left| \begin{array}{lllcl}
P^{J_{1}} & Q^{J_{1}J_{2}} &   Q^{J_{1}J_{2}J_{3}} & {\ldots} &
Q^{J_{1} \ldots J_{L}} \\
\rho^{\alpha} {\bar \rho}^{\ald} k^{J_{1}}_{\alpha \ald} &
P^{J_{2}} &  Q^{J_{2}J_{3}} & {\ldots} & Q^{J_{2} \ldots J_{L}} \\
0 & \rho^{\alpha} {\bar \rho}^{\ald} k^{J_{1}J_{2}}_{\alpha \ald} &
P^{J_{3}} & {\ldots} &  Q^{J_{3} \ldots J_{L}} \\
{\vdots} & {\vdots} & {\vdots} & {\vdots} & {\vdots} \\
0 & 0 & 0 & {\ldots} & P^{J_{L}} \end{array} \right| \]
  
Each entry of the $L \times L$ matrix in the above equation is a homogeneous
polynomial on the $CP^{1} \times CP^{1}$ space of degree $(1,1)$.
$Q$'s in the upper triangular part of the matrix represent the 
freedom which is to be fixed by  Eqs.(\ref{5.15i}) (or  Eqs.(\ref{5.19})).
Unfortunately, we have not been able to implement these
equations to fix this freedom.\\

{\bf Acknowledgments}\\
I benefited a lot from discussions with A.Rosly
whom I am very much obliged to.
This work was supported by INTAS grant 97-0103.

\end{document}